\documentclass[12pt]{article}
\usepackage{epsfig}
\def\br(#1,#2){\left\langle#1#2\right\rangle}
\def\sq(#1,#2){\left[#1#2\right]}
\def\s(#1,#2){s_{#1 #2}}
\def\t(#1,#2,#3){s_{#1 #2 #3}}

\begin{document}
\begin{titlepage}

\hspace*{\fill}\parbox[t]{5cm}
{hep-ph/0312024 \\
ANL-HEP-PR-03-104 \\
FERMILAB-Pub-03/378-T \\
RM3-TH/03-17 \\
ILL-(TH)-03-10 \\
NSF-KITP-03-98 \\ \\
\today} \vskip2cm
\begin{center}
{\Large \bf Associated Production of a $Z$ Boson and a \\
\bigskip  Single Heavy-Quark Jet} \\
\medskip
\bigskip\bigskip\bigskip\bigskip
{\large  {\bf J.~Campbell}$^1$,
         {\bf R.~K.~Ellis}$^2$,
         {\bf F.~Maltoni}$^3$,
     and {\bf S.~Willenbrock}$^{4,5}$} \\
\bigskip\bigskip\medskip
$^{1}$High Energy Physics Division,
Argonne National Laboratory \\
Argonne, IL\ \ 60439 \\ \bigskip
$^{2}$Theoretical Physics Department, Fermi National Accelerator Laboratory \\
P.~O.~Box 500, Batavia, IL\ \ 60510 \\ \bigskip
$^{3}$Centro Studi e Ricerche ``Enrico Fermi'' \\
via Panisperna 89/A, 00184 Rome, Italy \footnote{Mail address: Dipartimento di
Fisica, Terza Universit\`a di Roma, via della Vasca Navale 84, 00146 Rome,
Italy}\\ \bigskip
$^{4}$Department of Physics, University of Illinois at Urbana-Champaign \\
1110 West Green Street, Urbana, IL\ \ 61801 \\ \bigskip
$^{5}$Kavli Institute for Theoretical Physics, University of California \\
Santa Barbara, CA\ \ 93106 \\ \bigskip
\end{center}

\bigskip\bigskip\bigskip

\begin{abstract}
The leading-order process for the production of a $Z$ boson and a heavy-quark
jet at hadron colliders is $gQ\to ZQ$ ($Q=c,b$).  We calculate this cross
section at next-to-leading order at the Tevatron and the LHC, and compare it
with other sources of $ZQ$ events. This process is a background to new
physics, and can be used to measure the heavy-quark distribution function.
\end{abstract}

\end{titlepage}

\section{Introduction}
\label{sec:intro}

Many signals for new physics at hadron colliders involve electroweak gauge
bosons ($W^\pm,Z,\gamma$) and jets containing heavy quarks ($c,b$).  The prime
example is $W+4$ jets, with one or more jets containing a $b$ tag, which led to
the discovery of the top quark \cite{Abe:1995hr,Abachi:1995iq,Acosta:2001ct}.
Other examples include signals for the Higgs boson and the superpartners of
the known particles \cite{Carena:2003yi,unknown:1999fr}.  It is therefore
crucial to understand the standard-model background from the production of
electroweak bosons and heavy-quark jets with good accuracy.

\begin{figure}[ht]
\begin{center}
\vspace*{0cm} \hspace*{0cm} \epsfxsize=10cm \epsfbox{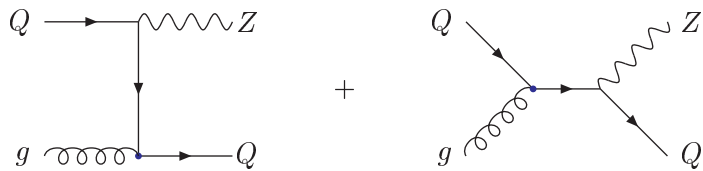}
\end{center}
\caption{Associated production of a $Z$ boson and a single high-$p_T$ heavy
quark ($Q=c,b$).} \label{fig:gQZQ}
\end{figure}

The simplest processes of this type are the production of a single electroweak
gauge boson and one heavy-quark jet.  In the case of the $W$ boson, the
leading-order process is $gs\to W^-c$ \cite{Baur:1993zd}, which has been
calculated at next-to-leading order (with $m_c$ nonzero) \cite{Giele:1995kr}.
For a photon, the leading-order process is $gQ\to \gamma Q$ ($Q=c,b$), which
has also been calculated at next-to-leading order (with $m_Q=0$)
\cite{Berger:1995qe,Bailey:1996px}. In this paper we consider the analogous
leading-order process for the $Z$ boson, $gQ\to ZQ$, shown in
Fig.~\ref{fig:gQZQ}, which we calculate at next-to-leading order (with
$m_Q=0$).

An alternative calculational scheme is to regard $gg\to ZQ\overline Q$ as the
leading-order process (with $m_Q$ nonzero), and to allow one heavy quark to be
emitted collinear to the beam, yielding a $ZQ$ final state. This approach has
two drawbacks. First, the expansion parameter of this calculation is
$\alpha_S\ln(M_Z/m_Q)$ rather than $\alpha_S$, so perturbation theory is less
convergent.  Using a heavy-quark distribution function sums these collinear
logarithms to all orders, resulting in a perturbative expansion in $\alpha_S$
and $1/\ln(M_Z/m_Q)$ \cite{Aivazis:1993pi,Collins:1998rz,Stelzer:1997ns}.
Second, it is much more difficult to obtain the next-to-leading-order
correction to $gg\to ZQ\overline Q$ than to $gQ\to ZQ$, both because there is
one more particle in the final state, and because the heavy-quark mass must be
maintained in the calculation of $gg\to ZQ\overline Q$ to regulate the
collinear region (one may set $m_Q=0$ in $gQ\to ZQ$, provided that the
heavy-quark transverse momentum is much larger than its mass). At present,
$gg\to ZQ\overline Q$ is known at next-to-leading order only for $m_Q=0$
\cite{Campbell:2000bg,Campbell:2002tg,Campbell:2003hd}.  In contrast, the
next-to-leading-order calculation of $gQ\to ZQ$ (with $m_Q=0$) that we perform
in this paper is sufficiently straightforward that the
next-to-next-to-leading-order calculation may be available in the foreseeable
future \cite{Glover:2002gz}.

The process $gQ\to ZQ$ is a background to $gb\to hb$, where the $Z$ boson and
the Higgs boson decay to the same final state ($b\bar b$, $\tau^+\tau^-$, or
$\mu^+\mu^-$) \cite{Choudhury:1998kr,Huang:1998vu,Campbell:2002zm}.  In
addition, $gQ\to ZQ$ could potentially be used to measure the $Q$ distribution
function. The $b$ distribution function is needed for the above-mentioned
Higgs-production process as well as for inclusive Higgs production, $b\bar
b\to h$ \cite{Dicus:1998hs,Balazs:1998sb,Maltoni:2003pn}. It is also needed
for the single-top production processes $qb\to q't$
\cite{Stelzer:1997ns,Stelzer:1998ni} and $gb\to W^-t$
\cite{Heinson:1996zm,Zhu:uj}, and the charged-Higgs production process $gb\to
H^-t$ \cite{Barger:1993th,Zhu:2001nt,Plehn:2002vy}.  The process $gQ\to \gamma
Q$ is much more sensitive to the charm distribution function than to that of
bottom, due to the greater electric charge of the charm quark
\cite{Abe:1998jc,Affolder:2001yp}.

At present, the $b$ distribution function is derived perturbatively from the
gluon distribution function
\cite{Aivazis:1993pi,Collins:1998rz,Pumplin:2002vw} and there is no direct
measurement of it.  The $c$ distribution function is similarly derived from
the gluon distribution function, and it agrees well with direct measurements.
Thus we expect the same to be true of the $b$ distribution function.  The
uncertainty in the $b$ distribution function derives from the uncertainty in
the gluon distribution function.

\begin{figure}[ht]
\begin{center}
\vspace*{0cm} \hspace*{0cm} \epsfxsize=10cm \epsfbox{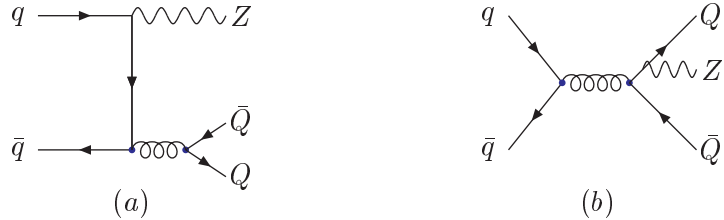}
\end{center}
\caption{Representative Feynman diagrams for $q\bar q\to ZQ\overline Q$. The
$Z$ boson may be radiated off (a) the initial-state quarks or (b) the
final-state quarks.} \label{fig:qqZQQ}
\end{figure}

Another source of $ZQ$ events is $q\bar q\to ZQ\overline Q$, shown in
Fig.~\ref{fig:qqZQQ}, where either one $Q$ is missed, or the two $Q$'s coalesce
into a single jet. We show that this process is much more significant at the
Tevatron than at the LHC, since light-quark distribution functions are
relatively more important at large values of Bjorken $x$.  This calculation is
carried out at leading order with a non-zero $Q$ mass.

\begin{figure}[ht]
\begin{center}
\vspace*{0cm} \hspace*{0cm} \epsfxsize=10cm \epsfbox{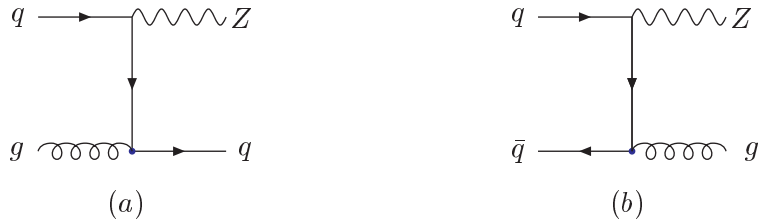}
\end{center}
\caption{Representative Feynman diagrams for $Zj$ production via (a) $gq\to
Zq$ and (b) $q\bar q\to Zg$.} \label{fig:Zj}
\end{figure}

In addition, we also calculate $Zj$ production at next-to-leading order, where
$j$ denotes a light-quark or gluon jet, as shown in Fig.~\ref{fig:Zj}
\cite{Arnold:1989dp,Gonsalves:1989ar,Giele:dj}. Using a silicon vertex detector
to tag heavy quarks, the probability that such a jet fakes a heavy-quark jet is
around $1\%$.  Taking this probability into account, we show that this source
of fake $ZQ$ events is comparable to genuine $ZQ$ events at the Tevatron, but
is relatively less important at the LHC.

All of the above processes can also lead to final states with two jets, with
varying numbers of heavy quarks.  For completeness, we also calculate these
cross sections (at leading order).  When combined with the
next-to-leading-order cross section for $ZQ$, one obtains the inclusive cross
section for $Z$ plus at least one heavy-quark jet (at next-to-leading order).

The next-to-leading-order calculations in this paper were performed with the
Monte-Carlo code MCFM \cite{Ellis:1999ec}.  The leading-order calculations
were performed both with this code and with MadEvent \cite{Maltoni:2002qb}.

\section{$gQ\to ZQ$ at NLO}
\label{sec:gQZQ}

The next-to-leading-order (NLO) calculation of $gQ\to ZQ$ parallels our NLO
calculation of $gb\to hb$ \cite{Campbell:2002zm}, and we refer the reader to
that work for a detailed discussion of the calculational scheme.  The
contributing subprocesses are
\begin{itemize}
\item $gQ\to ZQ$ (one loop)
\item $qQ\to ZQq$
\item $gQ\to ZQg$
\item $gg\to ZQ\overline Q$
\end{itemize}
The subprocess $q\bar q\to ZQ\overline Q$ is considered separately in the
following section, and is not regarded as a correction to $gQ\to ZQ$.

We work in the simplified ACOT scheme \cite{Aivazis:1993pi,Collins:1998rz},
which allows one to neglect the $Q$ mass throughout.  This is a good
approximation, and simplifies the calculation.  The error made by this
approximation is proportional to $1/\ln(M_Z/m_Q)\times m_Q^2/p_T^2$
\cite{Campbell:2002zm}.  We use the dipole-subtraction method
\cite{Ellis:1980wv} as formulated in Ref.~\cite{Catani:1996vz} to isolate and
subtract collinear divergences.

In our NLO calculation of $gQ\to ZQ$, we demand one and only one jet with
transverse momentum $p_T>15$ GeV within a rapidity range $|\eta|<2$ at the
Fermilab Tevatron.  This jet must contain a heavy quark.  At the CERN Large
Hadron Collider (LHC), the heavy-quark jet must have $p_T>15$ GeV and
$|\eta|<2.5$.

If two partons lie within a cone of radius $\Delta R<0.7$, we merge them into
a single jet with four-momentum equal to the sum of the two partons'
four-momenta.  This is done before the cuts described above are applied to the
jets.  The NLO process $gg\to ZQ\overline Q$ yields two heavy quarks in the
final state.  If they are merged into a single jet, we record it as a
double-heavy-quark jet.  This is only about 1\% of the $ZQ$ cross section.

Similarly, the NLO process $QQ'\to QQ'Z$, and processes related by crossing,
yield two heavy quarks in the final state.  However, these processes amount to
a correction of less than 1\%, so we neglect them \cite{Campbell:2002zm}.

We list in the first column of Tables~\ref{tab:tev} and \ref{tab:lhc} the LO
(in parentheses) and NLO cross sections for $gb\to Zb$ and $gc\to Zc$ at the
Tevatron and the LHC.  The $Zc$ cross section exceeds that of $Zb$ by 70\% at
the Tevatron and 35\% at the LHC because the charm distribution function is
larger than that of bottom. This is partially compensated by the fact that the
$Z$ has stronger coupling to bottom than to charm. The ratio of the $Zb$ and
$Zc$ partonic cross sections is proportional to
\begin{displaymath}
\frac{V_b^2+A_b^2}{V_c^2+A_c^2}=\frac{1+(1-{4\over 3}\sin^2\theta_W)^2}
{1+(1-{8\over 3}\sin^2\theta_W)^2}\approx 13/10\;.
\end{displaymath}
More importantly, a silicon vertex detector (SVX) can tag a $b$ jet with an
efficiency around 60\%, and a $c$ jet with an efficiency of about 15\%.  Thus
the majority of $ZQ$ events tagged with an SVX come from $Zb$.

The NLO processes that contribute to $ZQ$ also give rise to final states with
more than one jet, and we list these cross sections in the remaining columns.
These events are grouped in three classes.  The second column, labeled
$Z(Q\overline Q)$, corresponds to events with a single jet that contains two
heavy quarks.  As mentioned earlier, this is only about 1\% of the $ZQ$ cross
section.  The third column, labeled $ZQj$, corresponds to events with two
jets, one of which contains a heavy quark.  This is about $1/5$ of the $ZQ$
cross section at the Tevatron, and about $1/2$ at the LHC (for $p_T>15$ GeV).
The fourth column, labeled $ZQ\overline Q$, corresponds to events with two
jets, both of which contain heavy quarks. This is significantly less than
$ZQj$.  The $ZQ\overline Q$ events arise from the NLO process $gg\to
ZQ\overline Q$ (as do the $Z(Q\overline Q)$ events). The final column sums
these various processes, to give the inclusive cross section for $Z$ plus at
least one heavy-quark jet at next-to-leading order.

\begin{figure}[p]
\begin{center}
\hspace*{-.5cm}
\epsfig{figure=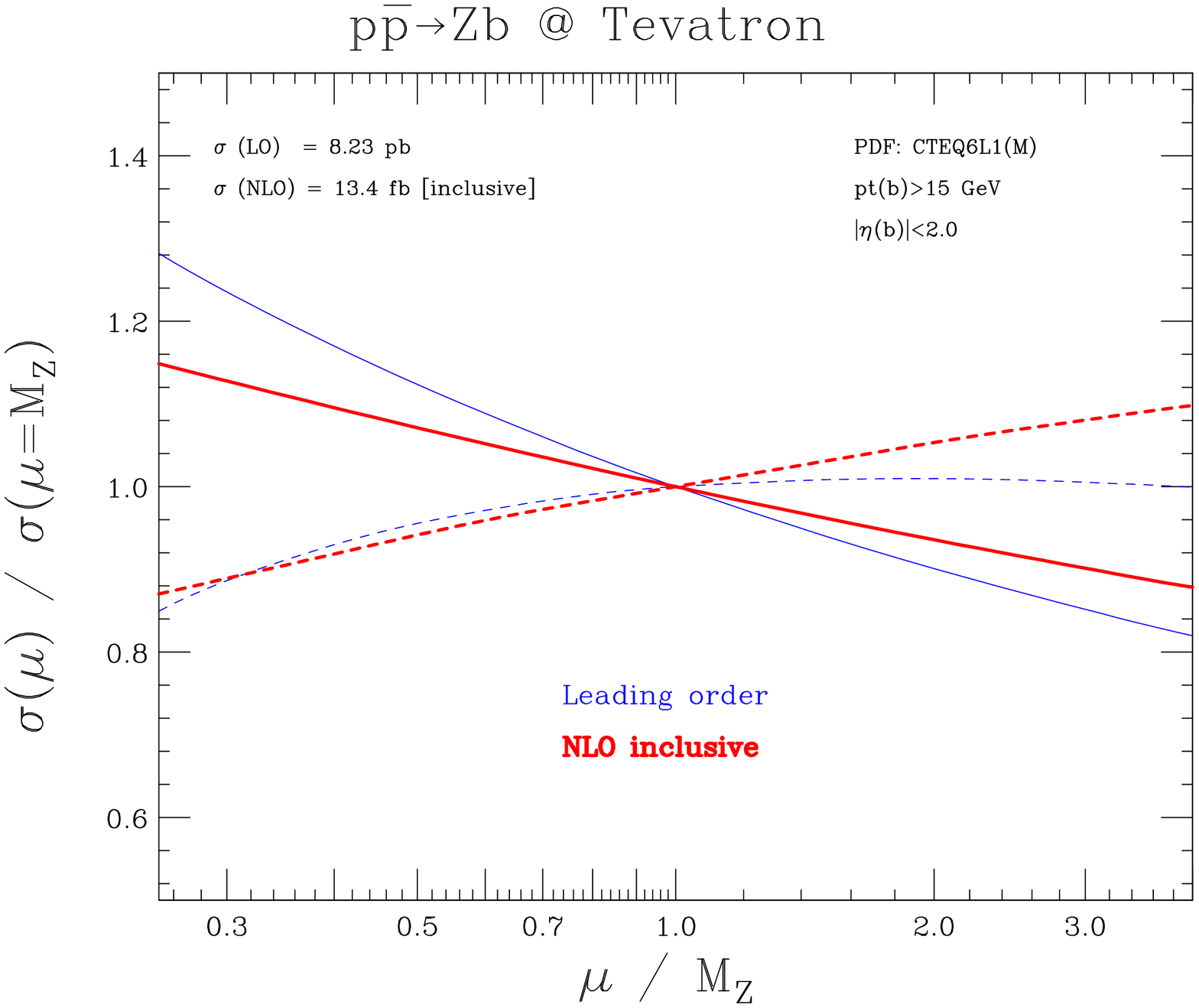,width=14cm,angle=90}
\vspace*{0cm} \caption{Cross section for $gb\to Zb$ at the Tevatron {\it vs}.\
the renormalization scale (solid curves) and factorization scale (dashed
curves). The ratio of the cross section at scale $\mu$ to the cross section at
scale $\mu=M_Z$ is plotted {\it vs}.\ the ratio of the scales.  The
next-to-leading-order (NLO) inclusive cross section (bold) is less sensitive to
the scales than the leading-order (LO) cross section (regular). }
\label{fig:mudep_tevb}
\end{center}
\end{figure}

\begin{figure}[p]
\begin{center}
\hspace*{-.5cm}
\epsfig{figure=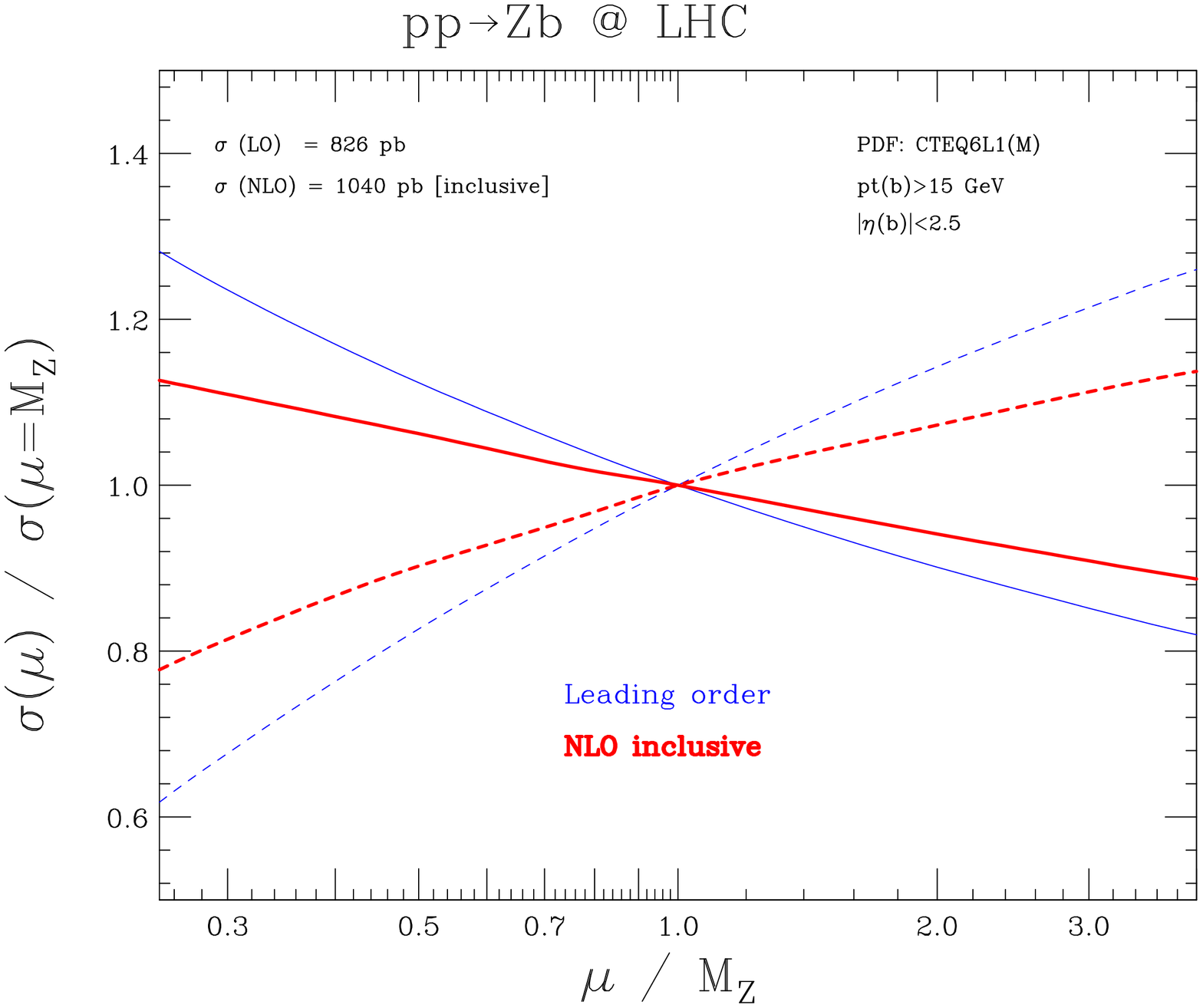,width=14cm,angle=90}
\vspace*{0cm} \caption{Same as Fig.~\ref{fig:mudep_tevb}, but at the LHC.}
\label{fig:mudep_lhcb_lolum}
\end{center}
\end{figure}

\begin{figure}[p]
\begin{center}
\hspace*{-.5cm}
\epsfig{figure=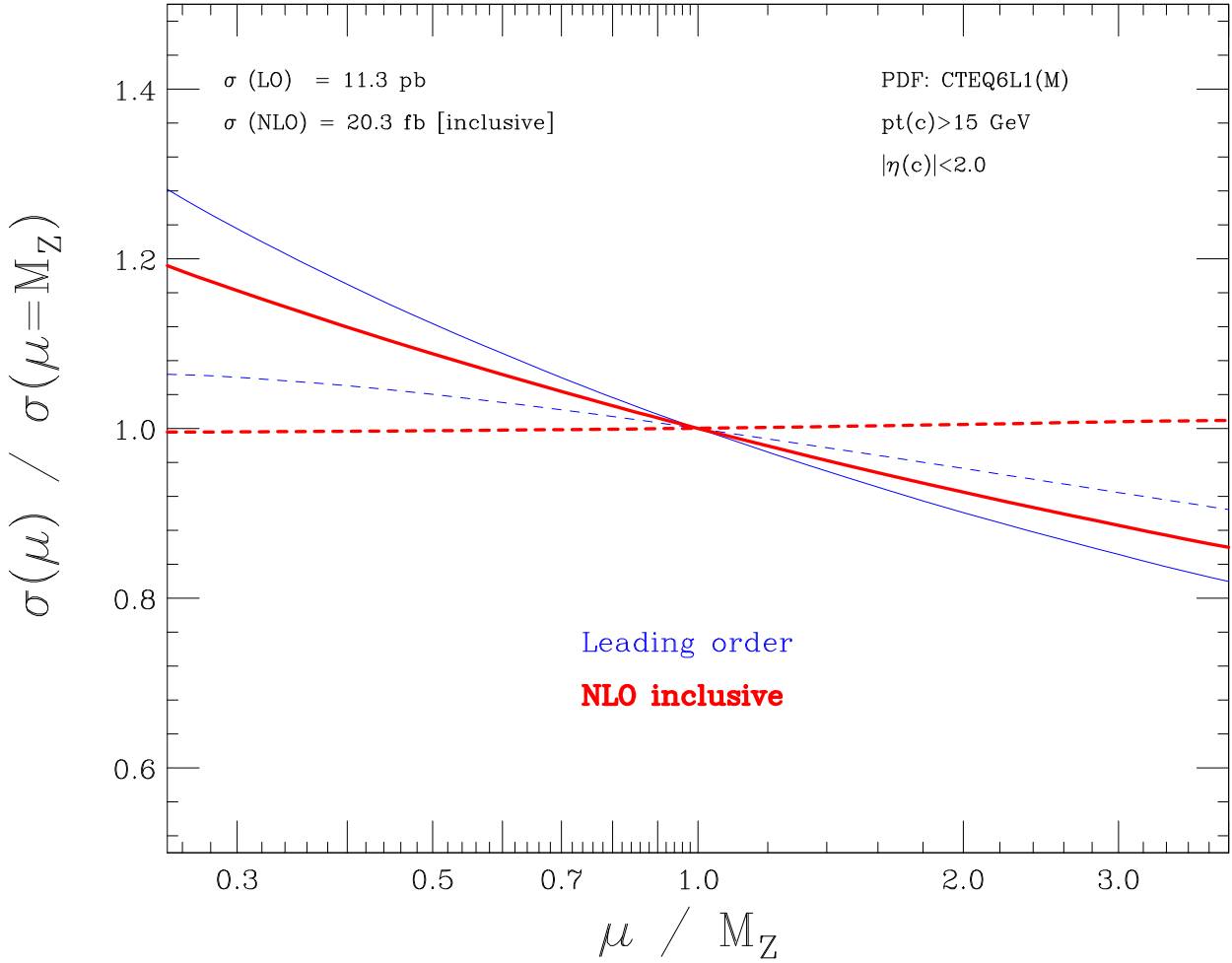,width=14cm,angle=90}
\vspace*{0cm} \caption{Same as Fig.~\ref{fig:mudep_tevb}, but for $Zc$
production.} \label{fig:mudep_tevc}
\end{center}
\end{figure}

\begin{figure}[p]
\begin{center}
\hspace*{-.5cm} \epsfig{figure=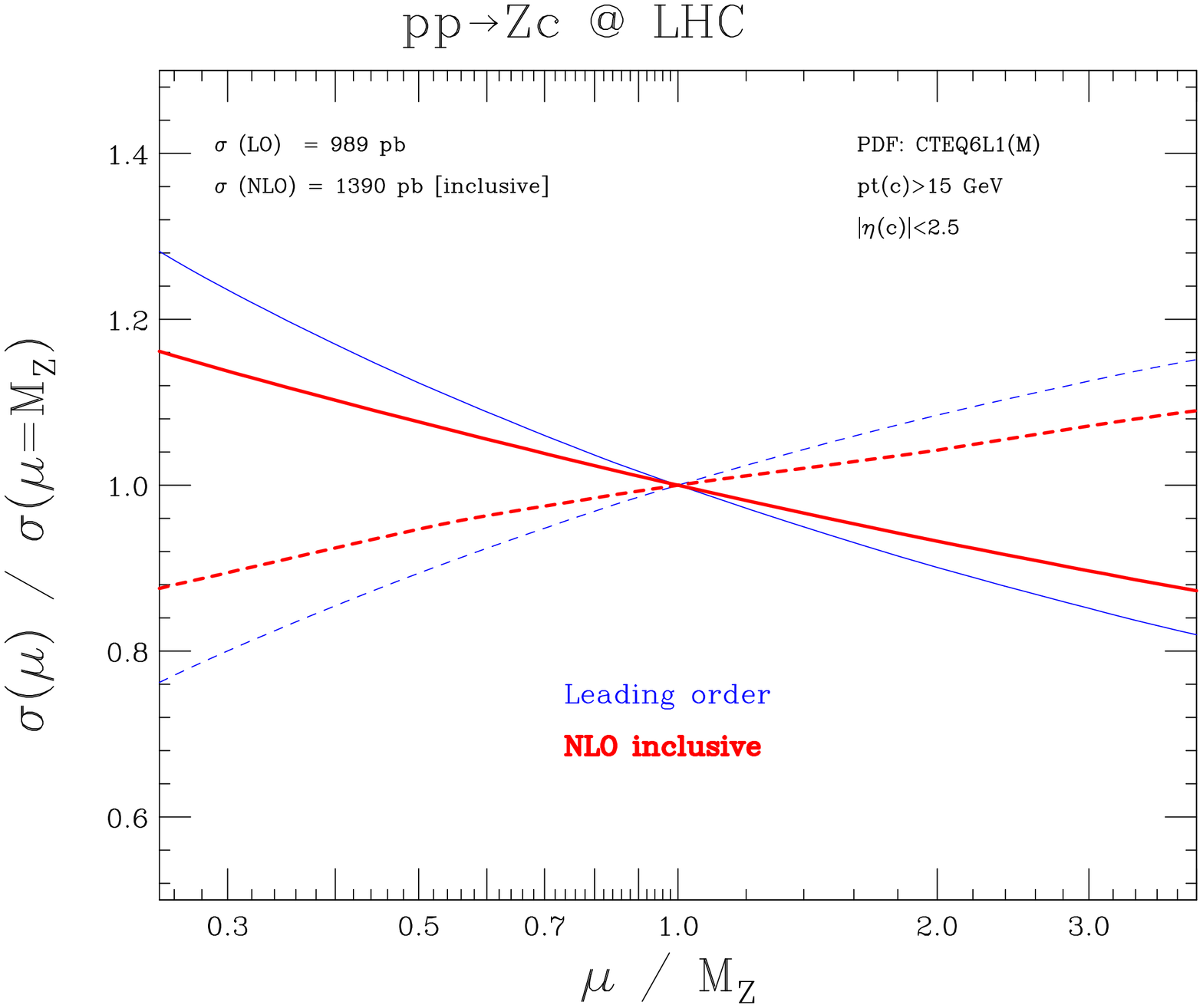,width=14cm,angle=90}
\vspace*{0cm} \caption{Same as Fig.~\ref{fig:mudep_tevc}, but at the LHC.}
\label{fig:mudep_lhcc_lolum}
\end{center}
\end{figure}

We show in Figs.~\ref{fig:mudep_tevb}--\ref{fig:mudep_lhcc_lolum} the scale
dependence of the inclusive cross section for $Zb$ and $Zc$ production at the
Tevatron and the LHC at leading order (LO) and next-to-leading order (NLO). The
dependence on the renormalization and factorization scales are shown
separately.  Both scale dependencies are reduced at NLO in comparison with
LO.  We use $\mu_R=M_Z$, $\mu_F=M_Z$ as our default value of the
renormalization and factorization scales.

We also estimate the uncertainty in the NLO inclusive cross section.  The
first uncertainty is due to varying the renormalization scale between half and
twice its default value of $\mu_R=M_Z$.  The second uncertainty, obtained in
the same manner, is due to the factorization scale.  The third uncertainty is
from the parton distribution functions \cite{Pumplin:2002vw}. There is also an
additional uncertainty of 4\% due to $\delta\alpha_S(M_Z)=0.002$.
\cite{Hagiwara:fs}

\begin{figure}[p]
\begin{center}
\epsfig{figure=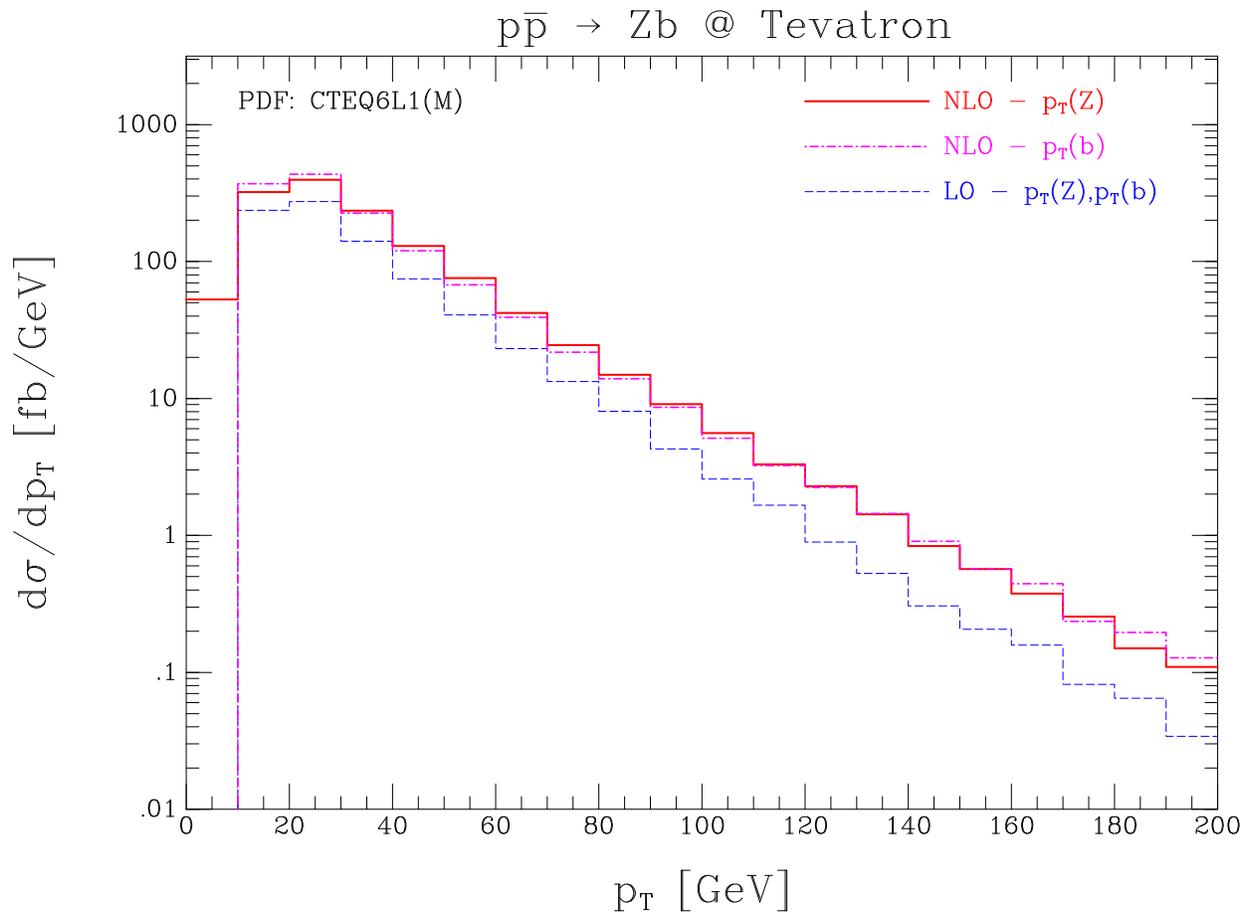,width=12cm,angle=-90}
\caption{The transverse-momentum distribution of $gb\to Zb$ at the Tevatron.
Qualitatively similar results are obtained for $gc\to Zc$.  The LO curve is
the $p_T$ of either the $Z$ or the $b$ jet. At NLO, the distributions of both
the $Z$ and the highest-$p_T$ $b$ jet are shown for the inclusive cross
section.} \label{fig:pt_b_tev}
\end{center}
\end{figure}

\begin{figure}[p]
\begin{center}
\epsfig{figure=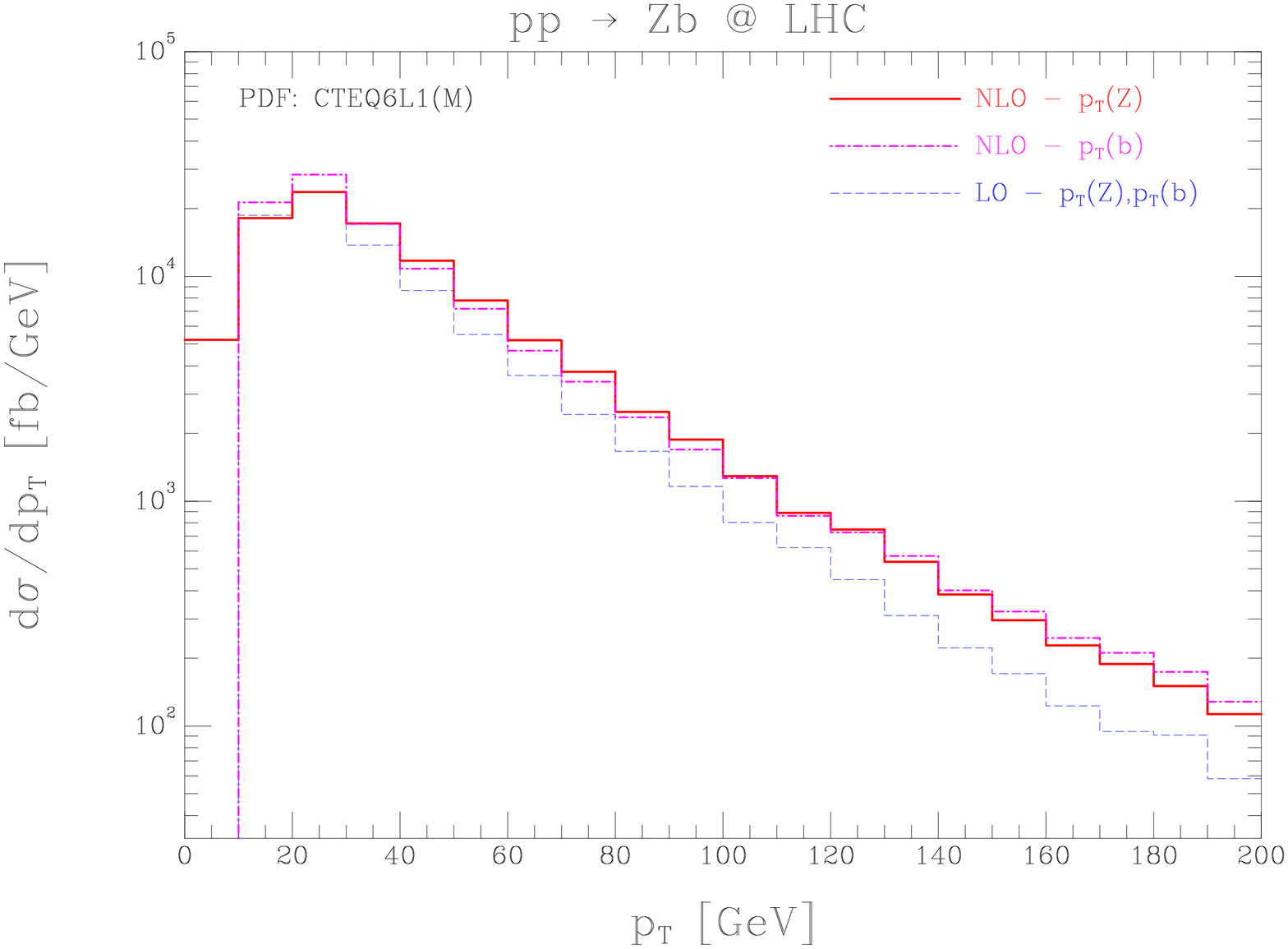,width=12cm,angle=-90}
\caption{Same as Fig.~\ref{fig:pt_b_tev}, but at the LHC.} \label{fig:pt_b_lhc}
\end{center}
\end{figure}

We show in Figs.~\ref{fig:pt_b_tev}--\ref{fig:pt_b_lhc} the transverse-momentum
distributions of both the $Z$ and the highest-$p_T$ $b$ jet at the Tevatron and
the LHC . At leading order these distributions are identical, since the $Z$
recoils against the $b$ jet. At next-to-leading order the distributions for
the inclusive cross section are slightly different, since the $Z$ can recoil
against two jets. The distributions for $Zc$ production are qualitatively
similar.


\begin{table}[t]
\caption{Cross sections (pb) for $Z$-boson production in association with
heavy-quark jets at the Tevatron ($\sqrt{s}=1.96$ TeV $p\bar p$).  A jet lies
in the range $p_T>15$ GeV and $|\eta|<2$.  Two final-state partons are merged
into a single jet if $R_{jj}<0.7$.  No branching ratios or tagging
efficiencies are included.  The labels on the columns have the following
meaning: $ZQ=$ exactly one jet, which contains a heavy quark; $Z(Q\overline
Q)=$ exactly one jet, which contains two heavy quarks; $ZQj=$ exactly two
jets, one of which contains a heavy quark; $ZQ\overline Q=$ exactly two jets,
both of which contain a heavy quark.  For the last set of processes, the
labels mean: $Zj=$ exactly one jet, which does not contain a heavy quark;
$Zjj=$ exactly two jets, neither of which contain a heavy quark.  For $ZQ$ and
$Zj$, both the leading-order (in parentheses) and next-to-leading-order cross
sections are given.  The last column is the next-to-leading-order inclusive
cross section, which is the sum of four previous columns.  The CTEQ6M parton
distribution functions are used throughout, except for the LO cross sections in
parentheses, where CTEQ6L1 is used \cite{Pumplin:2002vw}. The factorization
and renormalization scales are chosen as $\mu_F=\mu_R=M_Z$. The uncertainties
are from the variation of the renormalization scale, the factorization scale,
and the parton distribution functions, respectively.}

\addtolength{\arraycolsep}{0.2cm}
\renewcommand{\arraystretch}{1.5}
\medskip
\begin{center}
\begin{tabular}[4]{|c|cccc|c|}
\hline \hline \multicolumn{1}{|c|}{Cross sections (pb)} &
\multicolumn{5}{c|}{Tevatron}\\[1pt]
\hline
                           & $ZQ$         & $Z(Q\overline Q)$  &   $ZQj$  &  $ZQ\overline Q$  &
                           $ZQ$ inclusive
\\
\hline
 $g b \to Z b           $  & (8.23) 10.4& 0.169  &  2.19  &  0.631&
 $13.4\pm 0.9\pm 0.8\pm 0.8$
\\
 $ q\bar q \to Z b\bar b$  & 3.32       & 1.92   &  --    &  1.59 &   6.83
\\
\hline
 $g c \to Z c           $  & (11.3) 16.5& 0.130  & 3.22   & 0.49  &
$20.3\,^{+1.8}_{-1.5}\pm 0.1^{+1.3}_{-1.2}$ \\
 $ q\bar q \to Z c\bar c$  &   5.66     & 6.45   &  --    & 1.70  &
13.8\\
\hline & \multicolumn{2}{|c}{$Zj$} &  \multicolumn{2}{c|}{$Zjj$}     & $Zj$
inclusive
\\
\hline $ q\bar q \to Z g,
  gq      \to Z q $        & \multicolumn{2}{|c}{(876) 870}  &  \multicolumn{2}{c|}{ 137}
  &  $1010\,^{+44}_{-40}\,^{+9}_{-2}\,^{+7}_{-12}$
\\
\hline
\end{tabular}
\end{center}
\label{tab:tev}
\end{table}

\begin{table}[t]
\caption{Same as Table~\ref{tab:tev}, except at the LHC ($\sqrt{s}=14$ TeV
$pp$).  A jet lies in the range $p_T>15$ GeV and $|\eta|<2.5$.}
\addtolength{\arraycolsep}{0.2cm}
\renewcommand{\arraystretch}{1.5}
\medskip
\begin{center}
\begin{tabular}[4]{|c|cccc|c|}
\hline \hline \multicolumn{1}{|c|}{Cross sections (pb)} &
\multicolumn{5}{c|}{LHC}\\[1pt]
\hline
                           & $ZQ$    & $Z(Q\overline Q)$&  $ZQj$ & $ZQ\overline Q$ & $ZQ$ inclusive \\
\hline
 $g b \to Z b           $  &(826) 649&  11.3  &  304   &  78.1 & $1040\,^{+70}_{-60}\,^{+70}_{-100}\,^{+30}_{-50}$  \\
 $ q\bar q \to Z b\bar b$  &  24.3   &  13.5  &  --    & 11.4  & 49.2  \\
\hline
 $g c \to Z c           $  &(989) 921&   8.8  &  396   & 61.5  & $1390\pm 100^{+60}_{-70}\,^{+40}_{-80}$  \\
 $ q\bar q \to Z c\bar c$  &  36.7   &  41.7  &  --    & 11.3      & 89.7
\\
\hline
& \multicolumn{2}{|c}{$Zj$}    &  \multicolumn{2}{c|}{$Zjj$}     & $Zj$ inclusive\\
\hline $ q\bar q \to Z g,
  gq      \to Z q $        &
\multicolumn{2}{|c}{(13500) 11600} & \multicolumn{2}{c|}{4270}   &
$15870\,^{+900}_{-600}\,^{+60}_{-300}\,^{+300}_{-500}$
\\
\hline
\end{tabular}
\end{center}
\label{tab:lhc}
\end{table}


\section{$q\bar q\to ZQ\overline Q$ at LO}
\label{sec:qqZQQ}

Another contribution to $ZQ$ production comes from $q\bar q\to ZQ\overline Q$
($q=u,d,s$), shown in Fig.~\ref{fig:qqZQQ}, where one $Q$ is outside the
coverage of the detector. The dominant contribution to the cross section comes
from diagrams in which the $Z$ is radiated from the initial-state quarks while
the heavy quarks arise from gluon splitting, as shown in
Fig.~\ref{fig:qqZQQ}(a). We maintain the heavy-quark mass throughout the
calculation in order to regulate the singularity that would arise from a gluon
splitting to massless collinear quarks.  Although the NLO cross section for
this process with massless quarks is available
\cite{Campbell:2000bg,Campbell:2002tg,Campbell:2003hd}, the cross section with
massive quarks is known only at LO.

The LO cross section for $ZQ$ production via $q\bar q\to ZQ\overline Q$ is
given in the first column of Tables~\ref{tab:tev} and \ref{tab:lhc}.  At the
Tevatron, the cross section is about $1/3$ of $gQ\to ZQ$.  At the LHC, the
contribution from $q\bar q\to ZQ\overline Q$ is relatively much less, only
about $1/30$ of $gQ\to ZQ$.  This reflects the fact that this process is
initiated by a $q\bar q$ collision.  At the Tevatron, where the typical values
of Bjorken $x$ are relatively large, the valence quark distribution functions
are significant.  In contrast, the typical values of $x$ are relatively small
at the LHC, so the valence quark distribution functions are less important.

The second column in Tables~\ref{tab:tev} and \ref{tab:lhc} lists the
contribution from $q\bar q\to ZQ\overline Q$ when the $Q$ and $\overline Q$
merge into a single jet.  This is comparable in size to the contribution of
this process to $ZQ$ production.  This is due to the aforementioned enhancement
that arises when the heavy quarks are collinear.  This process is the dominant
source of events where the $Z$ is accompanied by a single jet that contains
two heavy quarks.

The last column gives the contribution of $q\bar q\to ZQ\overline Q$ when both
jets are within the coverage of the detector.  The $Zb\bar b$ cross section is
about 1/2 of the $Zb$ cross section from this process at both machines, and
the $Zc\bar c$ cross section is about 1/3 of the $Zc$ cross section.  This is
in contrast with the $gQ\to ZQ$ process, where the $ZQ\overline Q$ final state
(that arises at NLO) is much less than $ZQ$.

\section{$q\bar q\to Zg$, $gq\to Zq$ at NLO}
\label{sec:Zj}

A light-quark or gluon jet can fake a heavy-quark jet.  With a silicon vertex
detector, the mistagging rate is typically around 1\%. Since the cross section
for $Zj$ production is much greater than that for $ZQ$, it is potentially a
large source of fake $ZQ$ events.

We list in Tables~\ref{tab:tev} and \ref{tab:lhc} the LO (in parentheses) and
NLO cross sections for $Zj$ production.  The LO processes are $q\bar q\to Zg$
($q=u,d,s,c,b$) and $gq\to Zq$ ($q=u,d,s$), as shown in Fig.~\ref{fig:Zj}.  The
NLO processes also contribute to the $Zjj$ final state, and we list that cross
section in the Tables as well.  The $Zj$ cross section is almost two orders of
magnitude greater than that of $gQ\to ZQ$ at the Tevatron.  Taking into
account the 1\% mistagging rate, we see that $Zj$ is a significant source of
fake $ZQ$ events at the Tevatron.  In contrast, at the LHC the $Zj$ cross
section is not nearly as significant.  Thus there will be relatively fewer
mistagged events at the LHC.  Qualitatively similar results are obtained for
the ratio of $Zjj$ to $ZQj$.

\section{Conclusions}
\label{sec:conclusions}

The dominant contribution to $Zb$ production is $gb\to Zb$. At the Tevatron,
$q\bar q\to Zb\bar b$ also makes a significant contribution. Combining these
two contributions, the total inclusive cross section for $Zb$ production at the
Tevatron is about 20 pb.  Thus there are about 2000 inclusive $Zb$ events
produced within the coverage of the detector at the Tevatron for every 100
pb$^{-1}$ of integrated luminosity delivered. The cleanest decay mode of the
$Z$ boson is to leptons ($\ell=e,\mu$), with a branching ratio of 6.7\%.
Including a $b$-tagging efficiency of 60\% yields about 80 tagged $(Z\to
\ell^+\ell^-)b$ events for every 100 pb$^{-1}$.  There are slightly more tagged
events when one accounts for the fact that some events contain two heavy
quarks, either within the same jet or in separate jets.  With data sets
between 4000 and 8000 pb$^{-1}$ expected in Run II at the Tevatron, there will
be between 3200 and 6400 tagged $(Z\to \ell^+\ell^-)b$ events.

At the LHC, the contribution from $q\bar q\to Zb\bar b$ is much less
significant, so most of the cross section comes from $gb\to Zb$. The total
inclusive cross section for $Zb$ production is about 1090 pb, a factor of about
50 larger than at the Tevatron.  Thus there will be an enormous number of $Zb$
events at the LHC.

The total inclusive cross section for $Zc$ production at the Tevatron is about
70\% greater than that of $Zb$, while at the LHC, $Zc$ is about 35\% greater
than $Zb$. However, the SVX tagging efficiency for charm is about 1/4 of the
$b$-tagging efficiency, so there will be fewer tagged $Zc$ events than tagged
$Zb$ events.  In contrast, the number of $\gamma c$ events far outweighs the
number of $\gamma b$ events at both machines, since charm has twice the
electric charge of bottom and the charm distribution function is larger than
that of bottom.  Even taking the greater tagging efficiency of bottom into
account, the number of tagged $\gamma c$ events is larger than the number of
tagged $\gamma b$ events \cite{Abe:1998jc,Affolder:2001yp}.

$Zj$ events, where the jet is mistagged as a heavy quark, are a significant
source of fake $ZQ$ events at the Tevatron, but much less so at the LHC.  Thus
a larger fraction of the tagged $ZQ$ events at the LHC will be from genuine
heavy-quark production.  For this reason, and also due to the relatively
smaller contribution from $q\bar q\to ZQ\overline Q$, the LHC provides a
cleaner environment for the extraction of the heavy-quark distribution
functions via $gQ\to ZQ$.

In addition to the decay $Z\to \ell^+\ell^-$, there will be many $Z\to
\nu\bar\nu$ events ($BR=20\%$), which will yield heavy-quark monojets.  This
will also yield dijet events with large missing transverse momentum, with one
or both jets containing heavy quarks.

Regardless of the $Z$ decay mode, the majority of $Z+2$ jet events with a
single heavy-quark tag at the LHC come from $ZQj$, not $ZQ\overline Q$.  At the
Tevatron, where $q\bar q\to ZQ\overline Q$ is relatively more important, the
$ZQj$ and $ZQ\overline Q$ final states are comparable in size.

\section*{Acknowledgments}

\indent\indent We are grateful for conversations and correspondence with
L.~Christofek, T.~Junk, T.~Liss, K.~Pitts, D.~Stuart, M.~Whalley, and
J.~Womersley. F.~M. warmly thanks the Department of Physics of the ``Terza
Universit\`a di Roma'' for the kind hospitality and support. S.~W.~thanks the
Aspen Center for Physics for hospitality.  This work was supported in part by
the U.~S.~Department of Energy under contracts Nos.~DE-AC02-76CH03000 and
DE-FG02-91ER40677 and by the National Science Foundation under Grant
No.~PHY99-07949.



\end{document}